# Design of the LLRF control system for MA cavity at CSNS RCS

Yang Liu*, Xiang Li, Jian Wu

Institute of High Energy Physics, Beijing, China

Spallation Neutron Source Science Center, Dongguan, Guangdong, China

Abstract

The China Spallation Neutron Source (CSNS) beam power was successfully reached 125 kW with a low beam loss in February 2022. In order to increase beam power, during the summer in 2022, we employ magnetic-alloy (MA) cavity in the rapid cycling synchrotron (RCS). It is a wideband cavity (Q=2), allows the second harmonic rf (h= 4) operation, with the existing ferrite cavity to realize the dual-harmonic acceleration. The second harmonic (h=4) is used for the bunch shape control and alleviating the space charge effects. We design of the low-level RF(LLRF) control system for MA cavity, in this paper, we describe the system design and implementation, and the preliminary test results.

Ⅰ. Introduction

The China Spallation Neutron Source (CSNS) is a large-scale scientific facility for neutron science, and it also provides a powerful platform for multidisciplinary application research. it consists of an 80MeV $H^-$ linac, a 1.6-GeV rapid cycling synchrotron (RCS), a target station, linac to ring beam transport line and ring to target beam transport line[1]. The repetition frequency of RCS is 25 Hz, and the beam power was successfully reached 125 kW with a low beam loss in February 2022. The schematic layout of CSNS is shown in figure 1.

In order to increase beam power, during the summer in 2022, we employ magnetic-alloy (MA) cavity in the rapid cycling synchrotron (RCS). It is a wideband cavity (Q=2), allows the second harmonic rf (h= 4) operation, with the existing ferrite cavity to realize the dual-harmonic acceleration. The second harmonic (h=4) is used for the bunch shape control and alleviating the space charge effects. The main parameters of the RCS are listed in Table 1.

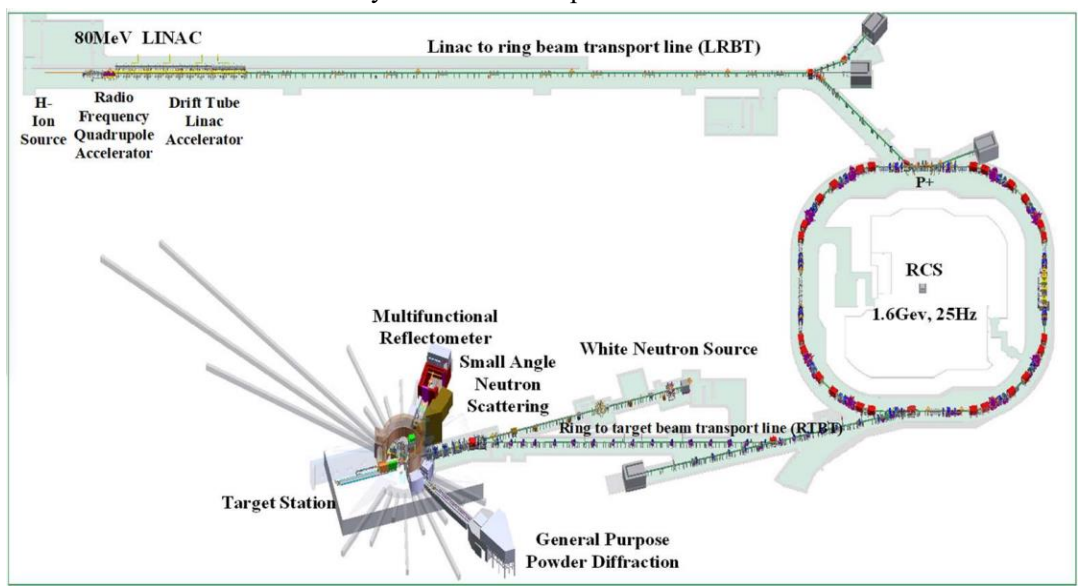

Fig. 1 schematic layout of CSNS

Table 1: The main parameters of RCS

| | |
|---|---|
| Circumference | 227.92 m |
| Energy | 0.08-1.6GeV |
| Repetition | 25Hz |
| Number of cavities | 8 ferrite cavities |
| | 1 MA cavity |
| Number of gaps | 2/3 |
| Fundamental RF frequency | 1.02-2.44MHz |
| Harmonic number | 2/4 |
| Maximum RF voltage | 180/50 kV |

Ⅱ．LLRF control system hardware

We employ CPCI bus platform for the LLRF control system. The hardware platform includes a CPCI6200 CPU board, a standard timing board, eight FPGA boards to realize eight ferrite cavities control, a FPGA board to realize magnetic-alloy cavity control, and four beam signal processing boards. The CPU board with EPICS IOC to realize control parameters setting and control status monitoring, the timing board to receive a global 25Hz timing signal and generation of event trigger, such as the extraction kicker magnets trigger pulse and the linac chopper pulse, the FPGA board to realize RF signal digitization data processing and control arithmetic for control function, the beam signal processing board to realize FCT, WCM, BPM data processing, a self-defined bus based on LVDS is adopted to transmit beam signal. A photograph of the FPGA board is shown in Fig.2.

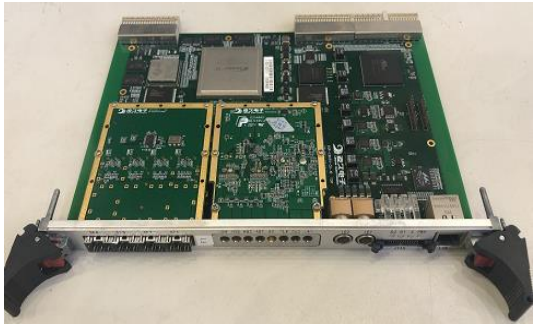

Fig. 2 Photograph of the FPGA board

The LLRF control system is a full-digital system, the FPGA board is the heart of the LLRF control system, it has a StratixIV FPGA (EP4SGX530) which provides 531.2K equivalent logic elements, four high-speed analog-to-digital converters (ADC, AD9268) and two digital-to-analog converters (DAC, AD9747), 8-bit digital I/O, SDRAM, optical interface, ethernet interface, and so on[2]. The schematic of FPGA board and LLRF control system interface is shown in Fig.3.

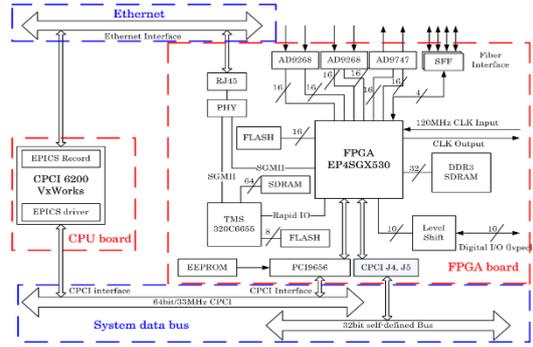

Fig. 3 The schematic of FPGA board and LLRF control system interface

Ⅲ．The multi-harmonic feedback control

The cavity gap voltage converted to pick up signal, pick up signal is directly digitized by ADCs operating at a clock frequency of 120 MHz. The sine and cosine signals for the I/Q demodulator and modulator are generated by a direct digital synthesizer(DDS). The phase pattern is obtained from the I/Q modulator to the demodulator, calibration the loop phase is necessary to I/Q control loop stability. The frequency signal (h=n) fed to the DDS is obtained by multiplying the revolution frequency (h=1) by the selected harmonic number. A 100 kHz bandwidth IIR low pass filter is used to obtained the in-phase and quadrature-phase (I/Q) values. The acquisition I/Q values is compared to the I/Q set pattern are I/Q error. I/Q error fed to PI controller and the I/Q modulator, the feedback output signal is generated. All the feedback signals are summed up to the RF driven signal, finally, the driving signal is generated by DACs. Fig.4 shows the block diagram of the multi-harmonic feedback control.

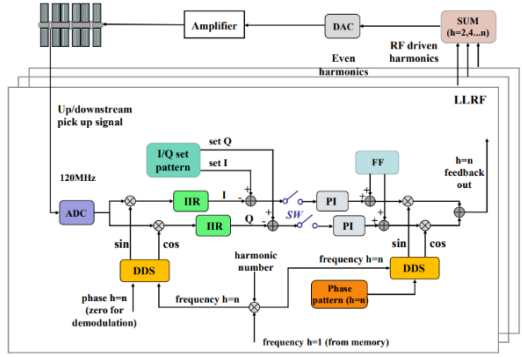

Fig. 4 The block diagram of the multi-harmonic feedback control

IV. Test results

The test for beam loading compensation was conducted during the summer in 2022, in the case with beam commissioning, without RF driving signal, acceleration of two bunches the maximum beam intensity is $1.95 \times 10^{13}$ ppp, only even harmonics are observed in the gap voltage, the maximum wake voltage values(h = 2,4,6,8) are about 10kV, 3kV, 1kV, 0.5kV. [5] Fig. 5 shows the wake voltage from injection to extraction.

When feedback control loop (h=2, 4, 6) are closed with the set points (Iset, Qset) set to (0, 0), all wake voltages are suppressed to less than 0.5kV, h=8 is less than 0.5 kV, so h=8 feedback control loop is not used. Fig. 6 shows the beam loading compensation with feedback control loops (h=2, 4, 6) closed.

With rf drive voltages was tested. In this case, MA cavity operation in the dual-harmonic mode, from injection to 2.1ms, MA cavity set the second harmonic RF voltage is 50kV, the fundamental RF voltage is 100kV by ferrite cavity, it is necessary because of the ratio of the harmonics. From 2.1ms to extraction, MA cavity provides the fundamental RF voltage is used to reduce the beam loss in the arc region. The voltage (h = 2,4) is controlled by the multi-harmonic feedback control loops. Fig.7 shows the harmonic components of the voltage without the beam when h=2, 4 feedback loop are closed.

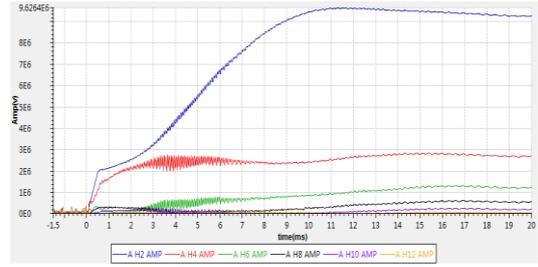

Fig. 5 The wake voltage in MA Cavity acceleration of two bunches at $1.95 \times 10^{13}$ ppp

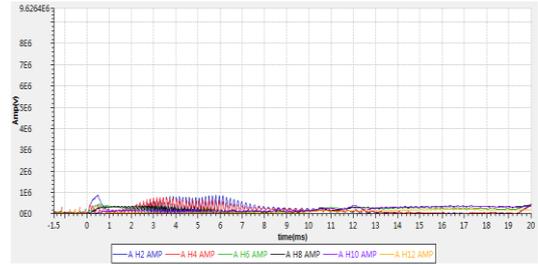

Fig. 6 The beam loading compensation with feedback control loops closed

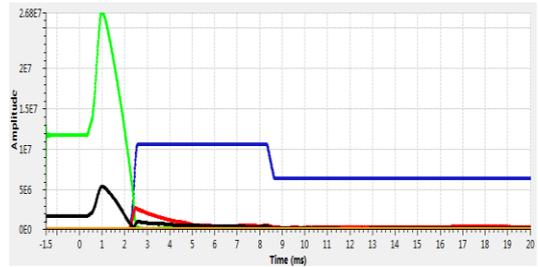

Fig. 7 The harmonic components of the voltage without the beam

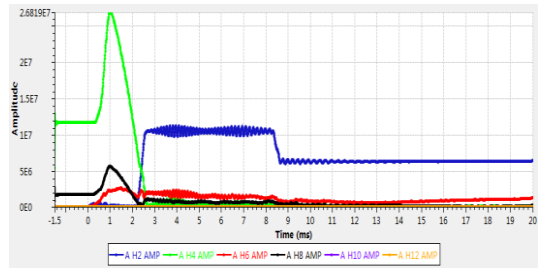

Fig. 8 The harmonic components of the voltage with an acceleration of a 140 kW beam

With the beam commissioning was also tested for dual-harmonic acceleration, in this case, eight ferrite cavities are providing the fundamental RF voltage, the maximum acceleration voltage of 180 kV. A MA cavity not only provides the second harmonic RF voltage but also provides the fundamental RF voltage. A single shot mode is used for the test, beam intensity is up to $1.95\times10^{13}$ ppp, corresponds to acceleration of a 140 kW beam at a repetition rate of 25 Hz[6]. Fig. 8 shows the harmonic components of the voltage with an acceleration of a 140 kW beam when feedback loop for (h=2, 4) are closed.

Ⅴ．Summary

We design of the LLRF control system for the wideband MA cavity, it is a full-digital system, the FPGA board is the heart. We describe the MA cavity multi-harmonic feedback control design, MA cavity is providing the second harmonic RF voltage, with the existing ferrite cavity to realize the dual-harmonic acceleration. Without beam was tested for the beam loading compensation, and with an acceleration of a 140 kW beam was also tested. The experimental results have confirmed the multi-harmonic feedback control works well.